\documentclass[journal]{IEEEtran}
\usepackage{fancyhdr}
\usepackage{amsmath,amssymb}
\usepackage{amsfonts}
\usepackage[dvips]{graphicx}
\usepackage{dsfont}
\usepackage{balance}
\usepackage{algpseudocode}
\usepackage{algorithm}

\usepackage[hidelinks]{hyperref}    %for using hyperlinks

\usepackage{color}
\usepackage{pgfplots}           %for plotting graphs
\pgfplotsset{compat=1.16}

\usepackage{enumerate}
\usepackage{graphics}
\usepackage{subfigure}
\usepackage{cite}
\usepackage{mathrsfs}
\usepackage{stfloats}

\usepackage{tikz}
\usepackage{mathdots}
\usepackage{yhmath}
\usepackage{cancel}
\usepackage{array}
\usepackage{multirow}
\usepackage{textcomp}
\usepackage{gensymb}
\usepackage{tabularx}
\usepackage{booktabs}
\usepackage{graphicx}

\usepackage{enumitem}
\newtheorem{theorem}{Theorem}

\newtheorem{proposition}{Proposition}
\newtheorem{remark}{Remark}

\begin{document}
	%\IEEEoverridecommandlockouts

\title{Energy-Aware Design of UAV-mounted RIS Networks for IoT Data Collection}
\author{
Dimitrios Tyrovolas,~\IEEEmembership{Student Member,~IEEE,}
Prodromos-Vasileios Mekikis,~\IEEEmembership{Member,~IEEE,}\\
Sotiris A. Tegos,~\IEEEmembership{Member,~IEEE,}
Panagiotis D. Diamantoulakis,~\IEEEmembership{Senior Member,~IEEE,}\\
Christos K. Liaskos,~\IEEEmembership{Member,~IEEE,} 
and George K. Karagiannidis,~\IEEEmembership{Fellow,~IEEE}
\thanks{ D. Tyrovolas, P.-V. Mekikis, S. A. Tegos, P. D. Diamantoulakis, and G. K. Karagiannidis are with the Wireless Communications and Information Processing (WCIP) Group, Electrical \& Computer Engineering Dept., Aristotle University of Thessaloniki, 54 124, Thessaloniki, Greece (e-mails: \{tyrovolas, vmekikis, tegosoti, padiaman,geokarag\}@auth.gr).
C. K. Liaskos is with the Computer Science Engineering Department, University of Ioannina,
45110 Ioannina, Greece (e-mail: cliaskos@cse.uoi.gr).}
\thanks{The research of D. Tyrovolas and P. D. Diamantoulakis has received funding from the European Union’s Horizon 2020 research and innovation programme under grant agreement No 957406. The research work of P.-V. Mekikis has received funding from the European Union´s Horizon 2020 research and innovation programme under Grant Agreement No. 891515. The research of S. A. Tegos and G. K. Karagiannidis has been co-financed by the European Regional Development Fund by the European Union and Greek national funds through the Operational Program Competitiveness, Entrepreneurship, and Innovation, under the call: “Special Actions: Aquaculture - Industrial Materials - Open Innovation in Culture (project code: T6YBP-00134). C. K. Liaskos acknowledges the support from H2020 COLLABS, EU871518.}
}
\maketitle

%\markboth{IEEE Transactions on Wireless Communications, SUBMITTED: Feb. 2009. }{M. Seyfi \MakeLowercase{\textit{et al.}}: On the Performance of  Selection Cooperation with Imperfect Channel Estimation}
%
%\markboth{ IEEE Transactions on Wireless Communications, AUG. 2011. }{M. Seyfi \lowercase{\textit{et al.}}: {Interference Imposed Relay Selection in the Underlay Cognitive Radio Networks}}

\begin{abstract} 
Data collection in massive Internet of Things networks requires novel and flexible methods.
Unmanned aerial vehicles (UAVs) are foreseen as a means to collect data rapidly even in remote areas without static telecommunication infrastructure. To this direction, UAV-mounted reconfigurable intelligent surfaces (RISs) aid in reducing the hardware requirements and signal processing complexity at the UAV side, while increasing the network's energy efficiency and reliability. Hence, in this paper, we propose the utilization of a UAV-mounted RIS for data collection and study the coverage probability in such networks. Additionally, we propose a novel medium access control protocol based on slotted ALOHA and Code Combining to handle the communication of multiple sensors. To account for the crucial energy issue in UAVs, we devise an energy model that considers both the UAV and the RIS weight, as well as the environmental conditions and the UAV's velocity. Finally, we characterize the performance of the proposed data collection scheme by analyzing the average throughput and the average data per flight, while providing useful insights for the design of such networks.

\end{abstract}

\begin{IEEEkeywords}
Reconfigurable Intelligent Surfaces (RIS), Unmanned Aerial Vehicles (UAVs), IoT Networking, Code Combining, Stochastic Geometry, Slotted ALOHA, Energy-awareness
\end{IEEEkeywords}

\IEEEpeerreviewmaketitle
%%%%%%%%%%%%%%%%%%%%%%%%%%%%%%%%%%%%%%%
\section{Introduction}

As a key enabler of smart cities, Internet of Things (IoT) networks will play a vital role in city monitoring by sensing the physical environment through a massive number of sensor nodes \cite{Guo}. Hence, it is imperative to extend the current capabilities of the sixth generation (6G) networks on ultra-massive IoT \cite{6G}. However, there exist two main constraints that prohibit a wider adoption, namely the maintainability and timely data collection. Specifically, the deploy-and-forget installation model should be followed, as maintaining each node individually is virtually impossible due to their extremely large numbers. Hence, energy-efficient communication becomes a crucial concern, given that the devices have to withstand for years using a single battery to allow a long-lasting operation \cite{eeiot}. At the same time, having ultra-low power transmitters with unknown locations affects the data collection, as the infrastructure should be excessively dense to guarantee service for all devices and, thus, impractical. To that end, novel and flexible methods of data collection should be proposed that are able to provide connectivity to a large area of randomly-deployed ultra-low power devices.

\subsection{State-of-the-Art}
Unmanned aerial vehicles (UAVs) are envisioned to play a pivotal role in the data collection for future IoT networks as they are able to satisfy the requirements for massive connectivity and increased throughput \cite{uavkarag,akis,wuUAV}. In more detail, considering their flexibility, UAVs can assist in the data collection from randomly-deployed sensors, while leveraging favorable characteristics of the established communication links in UAV-assisted networks. It should be mentioned, though, that the effective utilization of UAVs depends on their flight time duration, which is a function of their battery capacity as well as of the weight that they carry \cite{hossain1}. Specifically, the energy consumption of UAV-assisted data collection is crucial, since the UAVs have limited energy which is not only consumed for the communication process between the sensors and the access point (AP), but also for their movement \cite{eezhang}. Thus, it becomes of paramount importance to enhance the communication quality-of-service (QoS) in an energy-efficient way and utilize the available energy optimally.

% RIS can help eliminate the energy consumption for the telecommunication side due to their nearly passive nature
Recently, the concept of controllable wireless propagation through reconfigurable intelligent surfaces (RISs) has been introduced as a promising solution to significantly enhance the energy efficiency of future communication networks \cite{liaskos1,eealex}. In more detail, RISs have been introduced as programmable meta-surfaces, whose properties can be real-time altered and, thus, adjust to the network demands \cite{liaskos2}. Specifically, by altering the RISs' properties, a plethora of electromagnetic functions such as steering, diffusion, absorption, etc. can be implemented to the impinging signals, and thus by deploying RISs across the propagation environment, it is possible to convert it from an uncontrollable entity to an optimizable parameter, providing wireless connectivity seamlessly \cite{RRS}, \cite{casc}. Hence, a reliable and energy-efficient way to optimize the data-collection procedure for future IoT networks is to combine UAVs with RISs, which can facilitate the signal beamforming through their reflecting elements and, thus, enhance the network's reliability without increasing the sensors' transmission power, while keeping the UAV's power consumption low.

Over the last years, the synergy between UAVs and RISs has been studied extensively and it has been proven that it can improve significantly the performance of future wireless networks in many aspects \cite{direnzo1}. In more detail, considering the RISs' geometry, which enables the installment onto the facades of buildings or onto the UAVs, many works have studied the capabilities of:
\begin{itemize}
\item \textbf{Synergetic UAV-terrestrial RIS (TRIS) networks}: In synergetic UAV-TRIS networks, the RIS is attached to a fixed position, and has a favorable communication link with the UAV, offering enhanced QoS \cite{boulogeorgos}. In \cite{synergetic}, a synergetic UAV-TRIS communication system, combining a UAV with a highly directional antenna aiming at the TRIS was proposed and it was shown that it can optimize the network's reliability and average outage duration, which is of paramount importance for ultra-reliable and low latency communications (URLLC). Furthermore, different algorithms based on optimization theory or machine learning, which optimize jointly the UAV's trajectory and the TRIS passive beamforming have been proposed and proved that a synergetic UAV-TRIS network can enhance significantly the network's QoS in terms of achievable rate, secrecy rate, and blocklength \cite{uavdirenzo,secrecyuav,kaddoum}. Finally, in terms of IoT networking, in \cite{trungpoor}, a simultaneous wireless power transfer and information transmission scheme for IoT devices with support from a synergetic UAV-TRIS network was investigated, while, in \cite{timely}, a data collection framework assisted by a synergetic UAV-TRIS network has been presented.

\item \textbf{UAV-mounted RIS networks}: UAV-mounted RIS networks have been recently proposed as an interesting solution to maintain line-of-sight (LoS) links and enhance communication performance due to the offered additional degrees of freedom and flexible deployment. Specifically, in \cite{haas}, it was shown that a UAV-mounted RIS can offer enhanced outage performance whether the UAV-mounted RIS is moving or not. In addition, in \cite{trung}, the optimal UAV-mounted RIS deployment, as well as the optimal resource allocation to offer maximal reliability for a URLLC system with respect to the users’ fairness, were studied. Finally, \cite{chatzinotas} examined the performance in terms of reliability and spectral efficiency of a UAV-mounted RIS-assisted single-user network that co-exists with an ambient backscattering IoT system and proved that the deployment of a UAV-mounted RIS can offer enhanced performance for both systems.
\end{itemize}

\subsection{Motivation \& Contribution}

In the aforementioned works, it was shown that the exact location of the RIS affects the performance, since the RIS should be deployed in a specific orientation near to the AP or the ground users to maintain an optimal network performance \cite{bothsides}. However, in cases where the deployed sensors have time-variable QoS requirements, flexible deployment of the RIS that can be provided with the aid of UAVs can increase the ease of deployment and reduce the corresponding cost. Therefore, in this case, UAV-mounted RIS networks are a suitable option, since they are characterized by high deployment flexibility, 360-degree panoramic full-angle reflection, and favorable communication links due to the air-to-ground links' characteristics.

Most of the existing works show that by increasing the number of reflecting elements, the performance of the UAV-based network is enhanced in terms of coverage. It should be mentioned, though, that the increase of the RIS reflecting elements comes with the disadvantage of excess UAV weight and, thus, extra energy consumption leading to a decreased flight duration. However, to the best of the authors' knowledge, a UAV-mounted RIS-based data collection scheme that considers the flight duration has not yet been investigated in the existing literature. Moreover, to mitigate the communication performance drop from the decreased RIS size, alternative ways should be proposed, which will increase the average collected data. Hence, by taking into account the random-access nature of IoT data collection, an appropriate medium access control (MAC) protocol could improve the data collection procedure without using larger RIS or adding extra communication equipment.

To that end, in this paper, we propose a UAV-mounted RIS-based scheme for data collection for future IoT networks, and we analyze its performance in terms of reliability and average collected data. In more detail, our contribution is the following:
\begin{enumerate}[label=(\roman*)]
	\item We calculate the coverage probability of randomly-deployed sensors in a circular cluster using a UAV-mounted RIS that hovers above them, while considering imperfect phase estimation for the RIS due to UAV fluctuations.
	\item To make our model more realistic and account for the UAV energy restrictions, we propose an energy model that considers parameters such as the RIS operation and weight, the environmental conditions, and the UAV's kinematic condition. 
	\item To enhance the data-collection procedure, we propose a novel MAC-layer protocol that is based on slotted ALOHA and code combining (CC) and we analyze its performance in terms of coverage and average throughput.
	\item  To showcase the importance of the energy model, we propose a novel metric called \emph{average data per flight}, that provides the amount of data that can be collected from a UAV until it has to return for recharging purposes. 
\end{enumerate}

 %and  we propose the use of a novel metric, namely the \emph{average data per flight}, that takes into accoun

%(Xreiazetai na milisoume oti kanoume increase to coverage xari sto beamforming)

% Passive nature brings limitations regarding the range compared to active relays, but there are two solutions for this issue: i) Modified slotted ALOHA with code combining, and ii) increased RIS size
%(Allagi se modified slotted aloha me code combine)

%(Under the grant-free random access scheme, each active device directly transmits its data to the receiver without waiting for any permission)

%Nevertheless, the passive nature of RISs brings limitations regarding their communication range compared to active relays, but there are two solutions for this issue, i.e., i) the use of time diversity protocols, and ii) to increase the RIS size.
%Regarding the time diversity solution, HARQ-CC has been proven to be a way to achieve reliable data transmission by exploiting time diversity, but given that it is truncated to an L number of re-transmissions, it cannot guarantee better coverage at all cases. On the other hand, increasing the RIS elements can provide better coverage that comes with the disadvantage of excess UAV weight, and thus, extra energy consumption in the synergetic UAV-RIS scenario.

%To that end, there is a need to study the trade-off under which a UAV-RIS can provide a profitable and self-sustainable performance. The extra flexibility of UAVs provide substantial benefits to the data collection, but 

\subsection{Structure}\label{related}

The remaining of the paper is organized as follows. The system model is described in Section \ref{sysmod}. The performance analysis is given in Section \ref{analysis} and our results are presented in Section \ref{numres}. Finally, Section \ref{conc} concludes the paper.

%%%%%%%%%%%%%%%%%%%%%%%%%%%%%%%%%%%%%%%
\section{System Model}\label{sysmod}

We consider a set of $S$ uniformly-distributed single-antenna IoT sensors located inside a disk of radius $R$. Due to the low-maintenance requirements of such sensors and in order to expand their lifetime, they are transmitting with ultra-low-power. Therefore, due to the low transmit power as well as the density of the propagation environment, we assume that there is no direct link to serve the communication between each sensor and the AP.  To improve the received power at the AP, we employ a UAV-mounted RIS that is able to assist the communication by reflecting the sensor's transmissions towards the AP through a line-of-sight (LoS) link. Moreover, we assume that the UAV hovers at a height $h$ from the disk's center and the mounted RIS consists of $N$ reflecting elements. By taking into account the RIS reflection path, the baseband equivalent of the received symbol at the AP can be expressed as

\begin{equation}
    Y =  \sqrt{l_p G P_t} \sum_{i=1}^{N} \lvert H_{i1} \rvert \lvert H_{i2} \rvert e^{-j \left( \omega_i + \arg(H_{i1}) + \arg(H_{i2}) \right)}X + W,
\end{equation}
where $X$ is the transmitted signal for which it is assumed that $\mathbb{E}[|X|^2]=1$ with $\mathbb{E}[\cdot]$ and $\arg(\cdot)$ denoting expectation and the argument of a complex number, respectively \cite{RRS}. Also, $P_t$ denotes the sensor transmit power, $G = G_t G_r$ is the product of the sensor and the AP antenna gains,  and $H_{i1}$ and $H_{i2}$ are the complex channel coefficients that correspond to the $i$-th sensor-RIS and RIS-AP links, respectively. Moreover, $W$ is the additive white Gaussian noise with zero mean and variance $\sigma^2$, $\omega_i$ is the phase correction term induced by the $i$-th reflecting element, and $l$ is the path loss that corresponds to the sensor-RIS and RIS-AP links, respectively. Specifically, $l_p$ can be modeled as  
\begin{equation}
	l_p =   C_0 \frac{d_0^2}{\left(d_1 d_2\right)^{n}},
\end{equation}
where $n$ expresses the path loss exponent, $C_0$ denotes the product of the path loss of sensor-UAV and UAV-AP links at the reference distance $d_0$, while $d_1$ and $d_2$ denote the distances of the sensor-UAV and the UAV-AP links, respectively \cite{qingqing}. Considering the favorable characteristics of UAV communication links in cases where the UAV hovers in heights that allow LoS communication, the path-loss exponent can be assumed to be equal to $2$. Furthermore, it is assumed that there is no fading in the UAV-AP link considering the characteristics of air-to-air channels as it is assumed that the AP is located at the top of a building, e.g., UAV charging station, thus $\lvert H_{i2} \rvert = 1$ and $\arg(H_{i2})=\frac{2\pi r_i}{\lambda}$ with $\lambda$ and $r_i$ being the carrier's frequency wavelength and the distance between the UAV and the $i$-th reflecting element, respectively. Considering that the AP is located at the RIS's far-field, the distance $r_i$ is approximately equal to $d_2$. Additionally, it is assumed that $\lvert H_{i1} \rvert$ is a random variable (RV) following the Nakagami-$m$ distribution with shape parameter $m$ and spread parameter $\Omega$, which can describe accurately realistic communication scenarios characterized by severe or light fading. Finally, due to UAV fluctuations, each reflecting element does not adjust the phase perfectly to cancel the overall phase shift \cite{alouini1}. Thus, the received signal at the AP can be rewritten as
\begin{equation}
	Y =  \sqrt{l_p G P_t} H X + W,\label{yps}
\end{equation}
where $H =  \sum_{i=1}^{N} \lvert H_{i1} \rvert e^{-j \phi}$ and $\phi$ is an RV following the Von Mises distribution with concentration parameter $\kappa$ \cite{alouini1,justin}. By taking into account the results in \cite{justin}, $\frac{1}{N} \lvert H \rvert$ can be approximated by $\tilde{H}$, which is an RV following the Nakagami-$m$ distribution with shape parameter 
\begin{equation}
	\tilde{m} = \frac{ N \tilde{\Omega} {I_0}\left( \kappa \right)}{2 {I_0}\left( \kappa \right)+2 I_2\left( \kappa \right) -4 \tilde{\Omega} {I_0}\left( \kappa \right)},
\end{equation}
 and spread parameter 
 \begin{equation}
 	\tilde{\Omega} = \left(\frac{ {I_1}\left( \kappa \right)\Gamma \left(m + \frac{1}{2}\right) \sqrt{\Omega}}{ {I_0}\left( \kappa \right)\Gamma\left(m\right) \sqrt{m}} \right)^{2}, 
\end{equation}
where $I_p$ is the modified Bessel function of the first kind and order $p$ \cite{Mardia2009}.

\begin{figure}[!t]
	\centering
	\includegraphics[width=0.8\columnwidth]{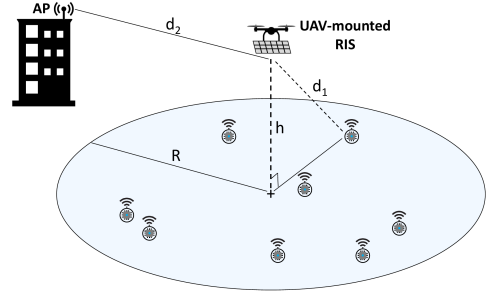}
	\caption{Network topology.}
	\label{sysmod_fig}
\end{figure}

Therefore, the instantaneous received SNR $\gamma_r$ of the proposed system can be expressed as 
\begin{equation}
	\gamma_r = \gamma_t  C_0 G N^2 \left(\frac{d_0}{d_1 d_2}\right)^{2} \tilde{H}^2,
\end{equation}
where $\gamma_t = \frac{P_t}{\sigma^2}$ is the transmit SNR and $\tilde{H}^{2}$ is a gamma-distributed RV with shape parameter $k = \tilde{m}$ and scale parameter $\theta= \frac{\tilde{\Omega}}{\tilde{m}}$. 
Considering that the sensors are distributed uniformly in a disc, we set $Z= d_1^{-2} \tilde{H}^{2}$, thus the instantaneous received SNR can be rewritten as
\begin{equation}
	\gamma_r = \gamma_t C_0 G N^2 Z \left(\frac{d_0}{d_2} \right)^{2}.
\end{equation}

\subsection{Slotted ALOHA-based Medium Access model}\label{aloha}

In the considered system, multiple IoT devices compete to access a shared AP through the UAV-RIS. Therefore, there is a need to enforce a methodology that handles access to the physical transmission medium \cite{SA}. We assume that the packets have fixed lengths and that transmissions are synchronized starting at the beginning of each slot. The slots have a duration defined by the time required for the transmission of one packet.

To enhance the performance of the proposed system, we propose the utilization of code combining (CC)\cite{alouiniCC}. Specifically, during any time slot, we assume that each device will be in any of three states: i) the idle state, when a device has no packet ready to transmit or if a new packet has just occurred while the terminal is waiting for the next time slot, ii) the transmission state, when a device transmits a packet (successfully or not), and iii) the retransmission state, when a device waits for retransmission in a future time slot after it has unsuccessfully tried to transmit a packet. More specifically, after a failed transmission attempt, the receiver stores the erroneous decoded frame and sends a NACK message. Once the frame is retransmitted, the new frame is combined with the stored one using maximum ratio combining (MRC) and the receiver tries to decode the combined frame. To maintain an acceptable latency, the truncated CC is considered, limiting the number of transmissions for a specific message to a maximum of $L$ \cite{alouiniCC}. Therefore, a device can be in the retransmission state for $L-1$ attempts, after which it updates its message and restarts the transmission process. It should be mentioned that, in the proposed communication system, it is assumed that no erroneous transmission of ACK and NACK messages exists.

\subsection{UAV Energy Model}\label{energymodel}
The lifetime of a UAV-RIS synergetic system depends on the battery capacity of the UAV $B_c$ at any given moment, as well as the total power consumption $P_{t}$ and can be calculated as

\begin{equation}
	L_t=\frac{B_c}{P_{t}},
\end{equation}
where the total consumed power $P_t$ is given by

\begin{equation}
	P_{t}= P_{\mathrm{thr}}+P_{\mathrm{tx/rx}} + P_{\mathrm{circ}}.
\end{equation}
Regarding $P_{\mathrm{tx/rx}}$, it is a flat cost for the battery provided for the navigational communication of the UAV and can be considered negligible ($\leq2$ Watt) compared to the drag counteract factor. Moreover, the term $ P_{\mathrm{circ}}$ refers to the consumed power due to the RIS circuitry that is responsible for its configuration and finally, the power for the UAV thrust, $P_{\mathrm{thr}}$, is a prevalent factor regarding the consumed energy, which includes all the power that is consumed for hovering, transiting, counteracting the wind drag, etc. Obviously, it is mainly affected by the weight and shape of the UAV and the additional carried components. To provide a more realistic model, we employ the state-of-the-art MN-505s KV320 motors from T-MOTOR. Based on the motors' datasheet \cite{motor}, the behavior of the consumed power for thrusting can be reliably characterized by the following equation:
\begin{equation}\label{pthr}
	P_{\mathrm{thr}}=4W^2+86W-21.2,
\end{equation}
where $W$ is given by 
\begin{equation}\label{w}
	W=U_w+B_w+R_w+S_w+D_w, 
\end{equation}
that includes all of the following weights, i.e., $U_w$, which is the weight of the UAV frame, while $B_w$ is the battery's weight and $R_w$ is the weight of the RIS given by $R_w=NE_w$, where $E_w$ is the weight of one reflecting element. Moreover, $S_w$ is the extra weight added to the motors due to any change in the speed of the UAV given by 
\begin{equation}
	S_w=(T_{\mathrm{max}}-U_w-R_w)\frac{S}{S_{\mathrm{max}}}, 
\end{equation}
where $T_{\mathrm{max}}$ is the maximum achievable thrust, $S$ is the average UAV speed, and $S_{\mathrm{max}}$ is the maximum achievable UAV speed. Finally, $D_w$ is the extra thrust needed by the motors to counteract the wind drag and it is given by
\begin{equation}
	D_w=\frac{\rho v_a^2 C_d A_{\mathrm{RIS}}}{2g},
%\frac{F_{\mathrm{dr}}}{g}
\end{equation}
where $\rho$ is the air density, $g$ is the gravity acceleration, $v_a$ is the average wind velocity, $C_d$ is the drag shape coefficient given experimentally by pre-calculated tables, and $A_{\mathrm{RIS}}$ is the area of the RIS side that is placed towards the airflow \cite{nature}. For the examined scenario, it is assumed that the RIS is placed in parallel with the ground and has a rectangular shape.

\section{Performance Analysis}\label{analysis}

In this section, we extract the analytical derivations that can be utilized to provide useful insights about the proposed network's coverage, as well as the average collected data for as long as the UAV-mounted RIS hovers in the sky. Specifically, we calculate four important metrics that characterize the performance of the considered network, namely i) the coverage probability of a sensor that is uniformly distributed in a disc, ii) the coverage probability of the same randomly-deployed sensor when CC is utilized, iii) the average throughput of the system when a slotted ALOHA-type medium access control (MAC) protocol is considered and finally, iv) the average data per flight, which takes into account the average throughput as well as the UAV's battery lifetime, which is affected by the RIS size and the environmental conditions.

\subsection{Coverage probability of a uniformly-distributed sensor}

Considering that the sensor is uniformly distributed inside a circular area with radius $R$ and the UAV is hovering above the disk's center at height $h$, the distance $d_1$ between the sensor and the UAV-RIS is an RV. Thus, the cumulative density function (CDF) of $d_1$ can be calculated through \cite{vpapanikk} and expressed as
    \begin{equation}
        F_{d_1}\left(x\right)= \frac{x^2-h^2}{R^2} , x\in \left[h, \sqrt{h^2 +R^2} \right].\label{ef}
    \end{equation}
Next, we provide an approximation of the coverage probability for the uniformly distributed sensor inside a circular area.
\begin{proposition}
	The coverage probability of a uniformly-distributed sensor can be approximated as
    \begin{equation}\label{11}
		\mathcal{P}_{c}  \approx  \frac{\theta}{R^2w} \sum_{i=0}^{\hat{k}-1} \frac{\gamma\left(i+1,\frac{h^2+R^2}{\theta}w\right)-\gamma\left(i+1,\frac{h^2}      {\theta}w\right)}{i!},
    \end{equation}
	where $i!$ is the factorial of $i$, $\gamma\left(\cdot\right)$ is the lower incomplete gamma function \cite{Coelho1998}, $\hat{k}$ is obtained by rounding $k$, $w= \frac{\gamma_{\mathrm{thr}} d_2^2}{\gamma_t d_0^2 C_0 N^2}$, and $\gamma_{\mathrm{thr}}$ is the received SNR threshold value.
\end{proposition}
\begin{IEEEproof}
The coverage probability of a uniformly-distributed sensor can be calculated through the probability of the complementary event, i.e., the outage probability. Specifically, the outage probability can be derived through the CDF of $Z$, which is given by
\begin{equation}\label{17}
	F_{Z}\left(x\right) = \int_{-\infty}^{\infty} F_{{\tilde{H}}^2} \left(\frac{x}{y}\right) f_{d_1^{-2}} \left( y \right) dy, %=\\
     %&=\int_{\frac{1}{h^2 + R^2}}^{\frac{1}{h^2}} \frac{\gamma\left(k,\frac{x}{\theta y}\right)}{\Gamma \left(k \right)} \frac{1}{\left(Ry\right)^{2}} dy ,
\end{equation}
where $F_{{\tilde{H}}^2}  = \frac{\gamma\left(k,\frac{x}{\theta y}\right)}{\Gamma \left(k \right)}$ is the CDF of ${\tilde{H}}^2$, $\Gamma \left(\cdot\right)$ is the gamma function \cite{Coelho1998}, and $f_{d_1^{-2}}$ is the probability density function (PDF) of RV $d_1^{-2}$, which equals to
\begin{equation}\label{pdf_apos}
    f_{d_1^{-2}} \left( x \right) = \frac{1}{{\left(xR\right)}^2}, x\in \left[\frac{1}{h^2 + R^2}, \frac{1}{R^2}\right].
\end{equation}
Using \eqref{pdf_apos}, \eqref{17} can be rewritten as
\begin{equation}\label{19}
      F_{Z}\left(x\right) = \int_{\frac{1}{h^2 + R^2}}^{\frac{1}{h^2}} \frac{\gamma\left(k,\frac{x}{\theta y}\right)}{\Gamma \left(k \right)} \frac{1}{\left(yR\right)^{2}} dy.
\end{equation}
By approximating k with $\hat{k}$, the lower incomplete gamma function is rewritten as 
\begin{equation}
    \gamma\left(k,\frac{x}{\theta y}\right) \approx \left( \hat{k} - 1 \right)! \left( 1 - e^{-\frac{x}{\theta y}} \sum_{i=0}^{\hat{k}-1} \frac{x^i}{\left(\theta y\right)^i i!} \right).
\end{equation}
After some algebraic manipulations, the CDF of $Z$ is derived. The coverage probability is defined as
\begin{flalign}\label{21}
\mathcal{P}_{c}  = 1- \Pr \left(\gamma_r \leq \gamma_{\mathrm{thr}} \right) = 1- \Pr \left( Z \leq w \right).
\end{flalign}
Considering \eqref{19} and \eqref{21}, the coverage probability can be calculated as in \eqref{11}, which concludes the proof.
\end{IEEEproof}

\subsection{Coverage Probability with Code Combining}

In order to enhance the performance of the proposed system, CC is utilized, as discussed in Section \ref{aloha}. Specifically, in order to maintain an acceptable latency, the \emph{truncated CC} is considered, limiting the number of retransmissions to a maximum of $L$ and, thus, the instantaneous received SNR after $l$ transmission rounds with CC with $l \in {1,...,L}$ is equal to

\begin{equation}
	\gamma_r^l = \gamma_t  C_0 G N^2 \left(\frac{d_0}{d_1 d_2}\right)^{2} \sum^{l}_{i=1} \tilde{H}_i^2,
\end{equation}
where $\tilde{H_i}$ is the channel at the $i$-th transmission round. It should be highlighted that the channels $\tilde{H}_i$ are assumed to be independent and identically distributed for every retransmission.
\begin{proposition}
	The coverage probability of the proposed system, i.e., the complementary event of the outage probability after $l$ transmission rounds with CC, is equal to
    \begin{equation}\label{prop2}
        \mathcal{P}_{c,l} =  \frac{\theta}{R^2 w}\sum^{ l \hat{k}-1}_{i=0}\frac{\gamma\left(i+1,\frac{h^2+R^2}{\theta}w\right)-\gamma\left(i+1,\frac{h^2}{\theta}w\right)}{i!}.
    \end{equation}
\end{proposition}

\begin{IEEEproof}
The coverage probability at the $l$-th CC round can be expressed as
\begin{equation}
    \mathcal{P}_{c,l} = \Pr\left(\gamma_r^l \geq \gamma_{\mathrm{thr}} \right) = 1- \Pr \left(\gamma_t G l_p N^2 \sum^{l}_{i=1} \tilde{H}_i^2 \leq \gamma_{\mathrm{thr}} \right).
\end{equation}
By invoking the moment matching technique, $S_2 = \sum^{l}_{i=1} \tilde{H}_i^2$ can be approximated by a gamma-distributed RV with shape parameter $\hat{k}_m = \frac{\mathbb{E}^{2}[S_2]}{\mathrm{Var}[S_2]}$ and scale parameter $\theta_m = \frac{\mathrm{Var}[S_2]}{\mathbb{E}[S_2]}$. Thus, in order to obtain the gamma distribution parameters, we need to calculate the first moment and the variance of $S_2$, which are equal to
\begin{equation}
	\mathbb{E}[S_2]= l \mathbb{E}[\tilde{H}^2] = l \tilde{\Omega},
\end{equation}
and
\begin{equation}
	\mathrm{Var}[S_2]= l \mathrm{Var}[\tilde{H}^2] = \frac{l{\tilde{\Omega}}^2}{\tilde{m}}.
\end{equation}
Thus, after the calculation of $\hat{k}_m$ and $\theta_m$ and following a similar procedure with the proof of Proposition 1, we obtain \eqref{prop2}, which concludes the proof.
\end{IEEEproof}

\subsection{Average Throughput}
At this point, we have investigated the conditions to achieve a successful decoding without considering that multiple sensors might access simultaneously the shared medium. Thus, we need to take into account the MAC protocol, as described in \ref{aloha}, to calculate the overall performance of the network by studying the average throughput. This is an important metric for data collection applications, which defines both the successful decoding capabilities of the proposed system and the efficiency of the utilized MAC protocol. Therefore, we provide the throughput analysis for the proposed IoT network from which we can gain useful insights into the scalability of the considered network. 

Considering that the derivation of the average throughput necessitates the calculation of the probability of successful decoding, which is affected by the utilized MAC protocol, we need to obtain the coverage probability at every transmission round.

\begin{theorem}
The coverage probability at the $l$-th CC round is given by \eqref{psucL} at the top of the next page, where $\hat{k}_b=\hat{k}$ and $\hat{k}_a= \left(l-1 \right)\hat{k}_b$.
\begin{figure*}
	\begin{equation}\label{psucL}
		\begin{split}
			& \mathcal{P}_{s,l} =\frac{{\theta}}{R^2w} \sum_{j=1}^{\hat{k}_b-1} \frac{\gamma\left(j+1,\frac{h^2+R^2}{{\theta}}w\right)-\gamma\left(j+1,\frac{h^2}      {{\theta}}w\right)}{j!} -\frac{{\theta}}{R^2w} \sum_{i=1}^{\hat{k}_a-1} \frac{\gamma\left(i+1,\frac{h^2+R^2}{{\theta}}w\right)-\gamma\left(i+1,\frac{h^2}{{\theta}}w\right)}{i!} \\
			& \quad + \frac{1}{\Gamma(\hat{k}_b) {{\theta}}^{\hat{k}_b}} \sum_{\mu=0}^{\hat{k}_a-1} \frac{1}{\mu!{{\theta}}^{\mu}} \sum_{\nu=0}^{\mu} \binom{\mu}{\nu}\frac{\left(-1\right)^{\mu-\nu} w^{\hat{k}_b+\mu}}{\left(\hat{k}_b+\mu -\nu \right) R^2} \left(\frac{\theta}{w}\right)^{\mu+k_b+1}\left[\gamma\left(\mu+\hat{k}_b+1,\frac{h^2+R^2}{{\theta}}w\right)-\gamma\left(\mu+\hat{k}_b+1,\frac{h^2}{{\theta}}w\right) \right].
		\end{split}
	\end{equation}
	\hrule
\end{figure*}
\end{theorem}
\begin{IEEEproof}
The probability of successful data reception at the $l$-th CC round is equal to the probability where $l-1$ transmissions were not adequate for successful decoding but invoking an extra transmission leads to successful data reception. Therefore it can be expressed as 
\begin{equation}
\mathcal{P}_{s,l} = \Pr\left(\gamma_r^{l-1} \leq \gamma_{\mathrm{thr}} \cap \gamma_r^{l} \geq \gamma_{\mathrm{thr}}  \right),
\end{equation}
which can be rewritten as
\begin{equation}
\mathcal{P}_{s,l} =\Pr\left( w d_1^{2} - \tilde{H}_l^2 \leq \sum^{l-1}_{i=1} \tilde{H}_i^2 \leq  w d_1^{2} \right) .
\end{equation}
Considering that $\tilde{H}_l^2$ and $ \sum^{l-1}_{i=1} \tilde{H}_i^2$ can be approximated as gamma distributed RVs with shape parameters $k_b$ and $k_a$ and scale parameters $\theta_a = \theta_b =\theta$, respectively according to Proposition 2, by conditioning on $d_1$, the above probability can be calculated as
\begin{equation}
\begin{split}
&\mathcal{P}_{s,l\vert d_1} =  \int_{0}^{ w d_1^{2}} \int_{w d_1^{2}-y}^{ w d_1^{2}}  f_b(y) f_a(x) dx dy  \\
& \qquad + \int_{w d_1^{2}}^{ \infty} f_b(y) dy \int_{0}^{w d_1^{2}} f_a(x) dx,
\end{split}
\end{equation}
where $f_v(x)$ with $v \in \{ a,b \}$ is the PDF of the gamma distribution with shape and scale parameters $k_v$ and $\theta_v$, respectively, given by
\begin{equation}
f_v(x) = \frac{1}{\Gamma(k_v) {\theta_v}^{k_v}} x^{k_v-1} e^{-\frac{x}{\theta_v}}, \ \forall x\in (0, \infty).
\end{equation}
After some algebraic manipulations, $P_{s,l}$ can be derived and expressed as

\begin{equation}\label{psuc}
\begin{split}
&\mathcal{P}_{s,l\vert d_1} = \frac{\gamma\left(k_a,\frac{w d_1^{2}}{\theta}\right)}{\Gamma \left(k_a \right)} -  \frac{\gamma\left(k_b,\frac{w d_1^{2}}{\theta}\right)}{\Gamma \left(k_b \right)} \\
& \quad + \frac{e^{-\frac{w d_1^{2}}{\theta}}}{\Gamma(\hat{k}_b) {{\theta}}^{\hat{k}_b}} \sum_{\mu=0}^{\hat{k}_a-1} \frac{1}{\mu!{{\theta}}^{\mu}} \sum_{\nu=0}^{\mu} \binom{\mu}{\nu}\frac{\left(-1\right)^{\mu-\nu} w^{\hat{k}_b+\mu}}{\left(\hat{k}_b+\mu -\nu \right)}.
\end{split}
\end{equation}
Finally, by taking into consideration the stochastic nature of $d_1$, we can decondition \eqref{psuc} on $d_1$ using \eqref{pdf_apos} and calculate the following expression
\begin{equation}\label{psuc2}
 \mathcal{P}_{s,l} =\int_{\frac{1}{h^2 + R^2}}^{ \frac{1}{h^2}} P_{s,l\vert y} \left(w\right)  f_{d_1^{-2}} \left( y \right) dy.
\end{equation}
To this end, substituting \eqref{psuc} into \eqref{psuc2}, \eqref{psucL} is derived, which concludes the proof.
\end{IEEEproof}
\begin{remark}
For $l=1$, the coverage probability at the $l$-th CC round $\mathcal{P}_{s,L}$ is equal with the coverage probability $\mathcal{P}_{c}$. 
\end{remark}

It is worth mentioning that for the case where CC is utilized, the number of transmissions can vary from one sensor to another depending on the channel conditions. Specifically, if the channel conditions are satisfactory, one transmission could be sufficient for error-free decoding. In the case of harsh channel conditions, $L$ CC rounds may be required to transmit successfully one data packet. Thus, in order to derive the average throughput for the proposed system, the average number of transmissions should be calculated.

\begin{proposition}
The average number of transmissions of an IoT network with $S$ sensors, which utilizes truncated CC with $L$ rounds is given by
\begin{equation}
\begin{split}
    &\Bar{T}_r = \sum^{L-1}_{i=1} i \sum^{i}_{j=1} \left[ 1 - \left(1-\rho\right)^{S-1}\right]^{i-j} \left(1-\rho\right)^{j(S-1)}\mathcal{P}_{s,j} \\
    &+L \Bigg[  \sum^{L-1}_{i=1} \left[ 1 - \left(1-\rho\right)^{S-1}\right]^{L-1-i} \left(1-\rho\right)^{i(S-1)}\left(1- \mathcal{P}_{c,i} \right)\\
 &+ \left[ 1 - \left(1-\rho\right)^{S-1}\right]^{L-1} \Bigg],
\end{split}
\end{equation}p
where $\rho$ is the access probability of a sensor, i.e., the probability to activate and transmit its data.
\end{proposition}
\begin{IEEEproof}
The mean value of the discrete random variable $T_r$ is given as
\begin{equation}
 \Bar{T}_r= \sum^{L}_{i=1} i \Pr(T_r=i) ,
 \end{equation}
where $\Pr(T_r=i) $ is the probability to finish the decoding procedure at the $i$-th transmission. Specifically, for the first round, the probability $\Pr(T_r=1) $ is equal to $(1-\rho)^{S-1} \mathcal{P}_{c}$, as in order for the procedure to be terminated at the first round, the transmitted data must be successfully decoded during the first transmission, i.e., a collision must not occur and the channel conditions must allow a successful message delivery. However, if $ i \in [2, L-1]$ we need to take into consideration all the combinations among the collision events and the channel conditions for each transmission. For instance, the transmitted data can be successfully decoded in the second round with probability $\left[ 1- \left(1-\rho\right)^{S-1}\right] \left(1-\rho\right)^{S-1} \mathcal{P}_{c} + \left(1-\rho\right)^{2\left(S-1\right)} \mathcal{P}_{s,2}$, meaning that two different events should be considered: i) the investigated sensor had a collision in the previous round, it transmits the second time without any collision from the other $(S-1)$ sensors and its message is successfully decoded, and ii) the sensor managed to transmit without collision in the first two rounds, its message was not decoded in the first round, but it is successfully decoded in the second round. It should be noticed that in the second event, the coverage probability is $\mathcal{P}_{s,2}$, as the receiver has already incomplete information about the transmitted message because of CC. Finally, in the last round, by taking into account that the transmission procedure will be terminated regardless of whether the message will be decoded or not, the probability $\Pr(T_r=L)$ is equal to the probability of unsuccessful message delivery due to collisions or unfavorable channel conditions at the previous to the last transmission round. Thus, by taking into account all the possible combinations among the collision events and the channel conditions for unsuccessful message delivery at the $L-1$ round, it stands
\begin{equation}
	\begin{split}
	&\Pr(T_r=L) =  \sum^{L-1}_{i=1} \left[ 1 - \left(1-\rho\right)^{S-1}\right]^{L-1-i} \left(1-\rho\right)^{i(S-1)} \\
	& \quad \times \left(1- \mathcal{P}_{c,i} \right)+ \left[ 1 - \left(1-\rho\right)^{S-1}\right]^{L-1}, 
	\end{split}
\end{equation}
which concludes the proof.
\end{IEEEproof}
\begin{remark}
For the case where CC is not utilized, the average number of transmissions $\Bar{T}_r =1$.
\end{remark}

After the calculation of the average number of transmissions, we present the system's average throughput for the considered scenario in the following proposition.
\begin{proposition}
The average throughput $\bar{R}$ of the proposed system can be calculated as
\begin{equation}
    \bar{R} =  \frac{B \mathrm{log}_2 \left(1 + \gamma_{\mathrm{thr}}\right)}{\bar{T}_r} \mathcal{P}_{suc}  ,
\end{equation}
where $B$ is the communication system's bandwidth and $P_{suc}$ is the probability of successful data decoding at any transmission attempt, which can be expressed as
\begin{equation}\label{psuc4}
\mathcal{P}_{suc} = \sum^{L}_{i=1} \sum^{i}_{j=1} \left[ 1 - \left(1-\rho\right)^{S-1}\right]^{i-j} \left(1-\rho\right)^{j(S-1)}\mathcal{P}_{s,j} .
\end{equation}
\end{proposition}

\begin{IEEEproof}
The average throughput of a communication system is equal to the rate multiplied by the probability of successful data reception $\mathcal{P}_{suc}$. Therefore, we need to calculate \eqref{psuc4}, which requires the inclusion of all possible combinations through which the transmitted message can be successfully decoded, i.e., in the first CC round if there is no collision, or in the second CC round after an unsuccessful decoding attempt in the first round, and so forth.  

To be more specific, for the first CC round, the probability of successful reception is given as $ {\left(1-\rho \right)}^{S-1} \mathcal{P}_c$ as, for the procedure to be successful at the first round, the transmitted data must be successfully decoded during the first transmission without any collision. Moreover, to calculate the successful decoding in the second CC round, we consider that the communication was unsuccessful in the first round due to collision or unfavorable channel conditions, while the second round is successful. Therefore, the probability of successful reception for the second CC round is given as $  \left[ 1 - \left(1-\rho\right)^{S-1}\right] {\left(1-\rho \right)}^{S-1} \mathcal{P}_c + {\left(1-\rho \right)}^{2(S-1)} \mathcal{P}_{s,2}$.

Similarly, for the following rounds, we need to take into consideration all the combinations among the collision events and the channel conditions for unsuccessful message delivery in the previous rounds, as well as the successful data reception at the ongoing round. Hence, by combining the results of each round, we arrive at the expression in \eqref{psuc4}. Finally, it should be noted that the average throughput is divided by $\bar{T}_r$ due to the fact that more than one transmission may be required in order to successfully decode the sensor's data, which concludes the proof.
\end{IEEEproof}

\begin{remark}
The probability of successful reception without CC is given as ${\mathcal{P}_{suc}}= \rho {\left(1-\rho \right)}^{S-1} \mathcal{P}_c$.
\end{remark}

\vspace{0.5 cm}
\subsection{Average Data per Flight} 
Although the throughput analysis is, in most cases, enough to study the data collection performance, the intrinsic requirements of a UAV-based system generate limitations and parameters that have to be taken into account. More specifically, the energy requirements of the UAV have to be considered in order to study reliably the data collection capabilities of the proposed synergetic UAV-RIS system. In this direction, we propose the use of a novel metric, namely the \emph{average data per flight}, that takes into account the throughput and various UAV parameters to calculate the amount of data that can be collected during the UAV lifetime, i.e., the total hovering time until the UAV has to return for recharging purposes. 
\begin{proposition}
	The average data per flight $\bar{D}_F$ is given by 
	\begin{equation}
		\bar{D}_F = \frac{B_c B \mathrm{log}_2 \left(1 + \gamma_{\mathrm{thr}}\right) }{\bar{T}_r \left[4W^2+86W-21.2+P_{\mathrm{tx/rx}} + P_{\mathrm{circ}}\right]} \mathcal{P}_{suc}.
	\end{equation}
\end{proposition}
\begin{IEEEproof}
	As discussed in Section \ref{energymodel}, the total consumed power $P_{t}$ includes the consumed power caused by thrust for hovering and counteracting the wind drag, the RIS circuitry consumption $P_{\mathrm{circ}}$, as well as $P_{\mathrm{tx/rx}}$ which is consumed for aviation purposes. In more detail, $P_{\mathrm{thr}}$ can be reliably characterized based on realistic equipment as in \eqref{pthr}, where $W$ is the total weight that the UAV is called to lift which is given in \eqref{w}. Thus, by using \eqref{pthr} and \eqref{w}, the average UAV flight duration is given by
	
	\begin{equation}
		L_t=\frac{B_c}{P_{t}}=\frac{B_c}{(4W^2+86W-21.2)+P_{\mathrm{tx/rx}}+P_{\mathrm{circ}}}
	\end{equation}
Therefore, by taking into consideration the average throughput of the proposed system, the average data per flight duration $\bar{D}$ can be calculated as
	\begin{equation}
		\bar{D}_F = {\bar{R}}{L_t},
	\end{equation}
which concludes the proof.
\end{IEEEproof}

\begin{table}[h!]
	\renewcommand{\arraystretch}{1.5}
	\caption{\textsc{Power Consumption Model Parameters}}
	\label{values}
	\centering
	\begin{tabular}{lll}
		\hline
		\bfseries Parameter & \bfseries Notation & \bfseries Value \\
		\hline\hline
		UAV weight				  &  $U_w$		& $3.25$ kg									\\
		RIS Element weight		  &  $E_w$		    & $ 7.66 \times 10^{-3}$ kg			        \\
		Battery weight			  &  $B_w$			& $1.35$ kg									\\
		Battery capacity          &  $B_c$  	    & $180$ Wh 		    						\\		
		%RIS Element number		  &  $N$ 			& $[100, 1000]$     						\\
             RIS circuitry consumption & $P_{\mathrm{circ}}$  & $0$ W                                    \\
		Communication required power   &  $P_{\mathrm{tx/rx}}$    & $1$ W 		    			\\
		Maximum achievable thrust &  $T_{\mathrm{max}}$ 		& $17$ kg		    			\\
		Maximum UAV speed		  &  $S_{\mathrm{max}}$		& $62$ km/h		  	  				\\		
		Air density				  &  $\rho$			& $1.225$ kg/m$^3$  						\\
		Air velocity			  &  $v_a$			& $2.5$ m/s (Light Air)  					\\
		Drag shape coefficient	  &  $C_d$    		& $1.28 @90^{\circ}$ or $0.005 @0^{\circ}$	\\
		RIS Area				  &  $A_{\mathrm{RIS}}$		& $N \frac{\lambda^2}{100}$ m$^2$ 	\\
		Gravity acceleration	  &  $g$			& 9.8 m/s$^2$								\\
		Motors					  &  Tmotor			& MN505-s KV320								\\
		\hline
	\end{tabular}
\end{table}

\section{Numerical Results}\label{numres}
In this section, the accuracy and validity of the derived expressions are verified through simulations. Furthermore, we provide insights related to the network's coverage, the performance of the proposed MAC protocol in terms of average throughput enhancement, as well as the network's data collection capabilities. In order to derive the numerical results, we set the parameters of the power consumption model as shown in Table \ref{values}. It should be mentioned that for the presented simulation results unless it is stated otherwise, the RIS is placed below the UAV and parallel to its bottom meaning that the drag shape coefficient $C_d$ is equal to 0.005. It is also assumed that the UAV is hovering above the ground without moving. Furthermore, the existence of light wind is assumed, which leads to random UAV fluctuations, while the area of each reflecting element is set equal to $\frac{\lambda^2}{100}$. Finally, the utilized simulation parameters are shown in Table \ref{values_sim}, where we assume that both the sensors and the AP are equipped with omnidirectional antennas and, thus, the gain parameter $G$ is set equal to 1. It should be highlighted that due to the existence of light wind, the concentration parameter $\kappa$ is set equal to 1, indicating the imperfect phase estimation due to random UAV fluctuations \cite{alouini1}.
\begin{table}[h!]
	\renewcommand{\arraystretch}{1.5}
	\caption{\textsc{ Simulation Results Parameters}}
	\label{values_sim}
	\centering
	\begin{tabular}{lll}
		\hline
		\bfseries Parameter & \bfseries Notation & \bfseries Value \\
		\hline\hline
		UAV height          &  $h$  	    & $50$ m 		    						\\
		Radius			  &  $R$			& $20$ m									\\
		Reference distance				  &  $d_0$		& $1$ m									\\
		Bandwidth				  &  $B$		& $125$ kHz 						\\
		%RIS Element number		  &  $N$ 			& $[100, 1000]$     						\\
		Transmit SNR   &  $\gamma_{t}$    & $95$ dB 		    						\\
		%Phase resolution power    &  $P_r$		    & $78$ mW 			    					\\
		SNR threshold	  &  $\gamma_{\mathrm{thr}}$		    & $0$ dB			    					\\
		Antenna Gain &  $G$ 		& $0$ dB 		    						\\
		Path Loss @ Reference distance		  &  $C_0$		& $-60$ dB		  	  						\\
		Shape Parameter			  &  $m$			& $3$ 						\\
		Spread Parameter			  &  $\Omega$			& $1$  						\\
		Concentration Parameter	  &  $\kappa$    		& $1$	\\
		%Gravity acceleration	  &  $g$			& 9.8m/s$^2$								\\
		%Propellers				  &  DJI			& 20 inch/6 inch							\\
		%Motors					  &  Tmotor			& MN505-s KV320								\\
		%Flight controller		  &  FC				& 120 A										\\
		\hline
	\end{tabular}
\end{table}

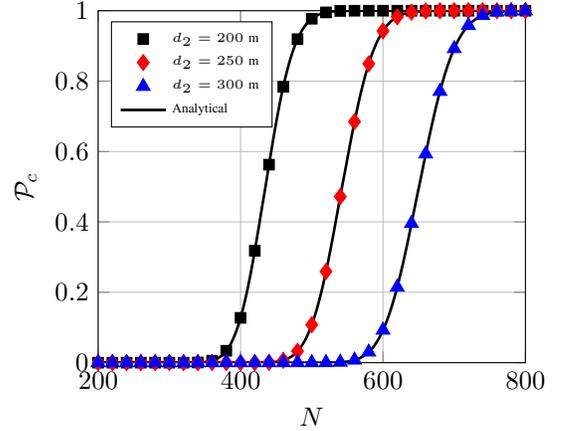
\begin{figure}
	\centering
	\begin{tikzpicture}
	\begin{axis}[
	width=0.82\linewidth,
	xlabel = {$N$},
	ylabel = {$\mathcal{P}_{c}$},
	xmin = 200,xmax = 800,
	ymin = 0,
	ymax = 1,
	xtick = {0,200,...,800},
	grid = major,
      legend style = {font = \tiny},
	legend cell align = {left},
	legend pos = north west
	]
	\addplot[
	black,
      only marks,
	mark=square*,
	mark repeat = 20,
	mark size = 2,
	]
	table {data_new/pcov1_200.dat};
	\addlegendentry{$d_2 = 200$ m}
	\addplot[
	red,
      only marks,
	mark=diamond*,
	mark repeat = 20,
	mark size = 3,
	]
	table {data_new/pcov1_250.dat};
	\addlegendentry{$d_2 = 250$ m}
	\addplot[
	blue,
      only marks,
	mark=triangle*,
	mark repeat = 20,
	mark size = 3,
	]
	table {data_new/pcov1_300.dat};
	\addlegendentry{$d_2 = 300$ m}
	\addplot[
	black,
       no marks,
	line width = 1pt,
	style = solid,
	]
	table {data_new/pcov1_200.dat};
	\addlegendentry{Analytical}
	\addplot[
	black,
      no marks,
	line width = 1pt,
	style = solid,
	]
	table {data_new/pcov1_250.dat};
	\addplot[
	black,
      no marks,
	line width = 1pt,
	style = solid,
	]
	table {data_new/pcov1_300.dat};
	\end{axis}
	\end{tikzpicture}
	\caption{Coverage probability versus $N$ for different UAV-AP distances.}
	\label{fig:coverage_1}
\end{figure}

In Fig. \ref{fig:coverage_1}, we illustrate the coverage probability $\mathcal{P}_{c}$ for the case of a uniformly distributed sensor versus the number of the RIS reflecting elements for three UAV-AP distances. For all distances, the simulation results validate our theoretical analysis by providing an exact match. Moreover, we observe that as the UAV-AP distance increases, the number of the reflecting elements should be increased to maintain an ultra-reliable data collection. For instance, as the UAV-AP distance increases by 100 m, about $100$ elements should be added to provide the same performance. Therefore, by taking into consideration the distance between the AP and UAV, each UAV must be equipped with an appropriate RIS to offer a specific coverage probability. However, for large distances, where a large number of reflecting elements is needed, the flight duration significantly decreases and, thus, the data collection performance deteriorates. Hence, it is of paramount importance to utilize alternative methods that improve the reliability performance, without increasing the number of the reflecting elements and, thus, the RIS weight.

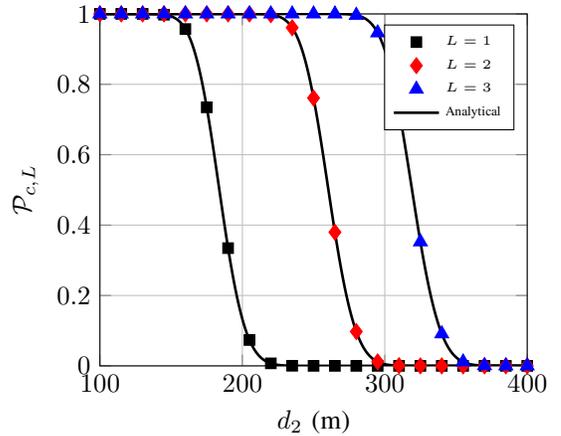
\begin{figure}
	\centering
	\begin{tikzpicture}
	\begin{axis}[
	width=0.82\linewidth,
	xlabel = {$d_2$ (m)},
	ylabel = {$\mathcal{P}_{c,L}$},
	xmin = 100,xmax = 400,
	ymin = 0,
	ymax = 1,
	xtick = {0,100,...,400},
	grid = major,
      legend style = {font = \tiny},
	legend cell align = {left},
	legend pos = north east
	]
	\addplot[
	black,
      only marks,
	mark=square*,
	mark repeat = 15,
	mark size = 2,
	]
	table {data_new/Pcov1.dat};
	\addlegendentry{$L=1$}
	\addplot[
	red,
      only marks,
	mark=diamond*,
	mark repeat = 15,
	mark size = 3,
	]
	table {data_new/Pcov2.dat};
	\addlegendentry{$L=2$ }
	\addplot[
	blue,
      only marks,
	mark=triangle*,
	mark repeat = 15,
	mark size = 3,
	]
	table {data_new/Pcov3.dat};
	\addlegendentry{$L=3$ }
	\addplot[
	black,
       no marks,
	line width = 1pt,
	style = solid,
	]
	table {data_new/Pcov1.dat};
	\addlegendentry{Analytical}
	\addplot[
	black,
      no marks,
	line width = 1pt,
	style = solid,
	]
	table {data_new/Pcov2.dat};
	\addplot[
	black,
      no marks,
	line width = 1pt,
	style = solid,
	]
	table {data_new/Pcov3.dat};
	\end{axis}
	\end{tikzpicture}
	\caption{Coverage probability with CC versus $d_2$ for $N=400$.}
	\label{fig:coverage_L}
\end{figure}

Fig. \ref{fig:coverage_L} depicts the effect of utilizing the truncated CC on the extension of the coverage probability for the case where the UAV-mounted RIS consists of 400 reflecting elements. As it can be observed, by invoking the truncated CC technique, the network's coverage can be efficiently expanded without increasing the number of the RIS reflecting elements. In more detail, by performing $L=3$ transmissions, the proposed system can offer high-reliability data collection for UAV-AP distances approximately equal to 300 m while, for the $L=1$ case, the system offers reliable data collection for distances less than or equal to 150m. Thus, the utilization of truncated CC can improve the network's coverage and enable reliable data collection for even greater distances from the AP, without increasing the number of the RIS reflecting elements, which can lead to increased RIS weight and, thus, increased power consumption.

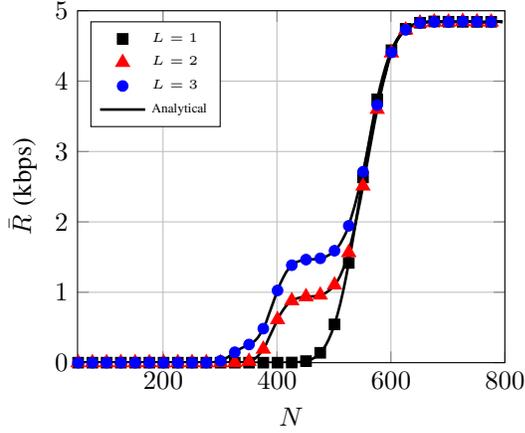
\begin{figure}
	\centering
	\begin{tikzpicture}
	\begin{axis}[
	width=0.82\linewidth,
	xlabel = {$N$},
	ylabel = {$\bar{R}$ (kbps)},
	xmin = 50,xmax = 800,
	ymin = 0,
	ymax = 5,
	ytick = {0,1,...,5},
	xtick = {0,200,...,800},
	grid = major,
      legend style = {font = \tiny},
	legend cell align = {left},
	legend pos = north west
	]
	\addplot[
	black,
      only marks,
	mark=square*,
	mark repeat = 5,
	mark size = 2,
	%line width = 1pt,
	%style = solid,
	]
	table {data_new/Throughput1_250.dat};
	\addlegendentry{$L=1$}
	\addplot[
	red,
      only marks,
	mark=triangle*,
	mark repeat = 5,
	mark size = 3,
	%line width = 1pt,
	%style = solid,
	]
	table {data_new/Throughput2_250.dat};
	\addlegendentry{$L=2$}
	\addplot[
	blue,
      only marks,
	mark=*,
	mark repeat = 5,
	mark size = 2,
	%line width = 1pt,
	%style = solid,
	]
	table {data_new/Throughput3_250.dat};
	\addlegendentry{$L=3$}
	\addplot[
	black,
      no marks,
	%mark=square,
	%mark repeat = 10,
	%mark size = 2,
	line width = 1pt,
	style = solid,
	]
	table {data_new/Throughput1_250.dat};
	\addplot[
	black,
      no marks,
	%mark=triangle,
	%mark repeat = 10,
	%mark size = 2,
	line width = 1pt,
	style = solid,
	]
	table {data_new/Throughput2_250.dat};
	%\addlegendentry{HARQ$^{I}$}
	\addplot[
	black,
      no marks,
	%mark=x,
	%mark repeat = 10,
	%mark size = 3,
	line width = 1pt,
	style = solid,
	]
	table {data_new/Throughput3_250.dat};
	\addlegendentry{Analytical}
\end{axis}
\end{tikzpicture}
	\caption{Average throughput versus $N$ for $d_2=250$ m.}
	\label{fig:Throughput}
\end{figure}

Next, Fig. \ref{fig:Throughput} portrays the average throughput without CC and for two CC cases, i.e., $L=2$ and $L=3$. Again, the simulation results validate the theoretical analysis. Furthermore, it can be observed that, for $N \in \left[300,500\right]$, the utilization of truncated CC enhances the system's average throughput and, thus, the proposed MAC protocol outperforms slotted ALOHA for the specific reflecting elements range. In addition, it can be noticed that the average throughput saturates in all cases when $N \geq 650$, as it reaches the maximum achievable rate due to the offered gain by the RIS. Increasing $N$ further than 650 would not only be unnecessary for the throughput, but it would also increase the UAV’s power consumption due to the extra RIS weight. Finally, it should be mentioned that the shape of the CC curves in the $N \in \left[300,500\right]$ range is caused by the fact that each consecutive retransmission leverages information from the previous one (i.e., due to the utilization of MRC). In detail, as $N$ decreases, the channel gain deteriorates and, thus, the proposed MAC protocol enables the sensors to initiate retransmission attempts to restore the communication. 
%Therefore, due to the different channel gain in every extra retransmission round, the average throughput for the proposed MAC protocol behaves as in Fig. \ref{fig:Throughput}.

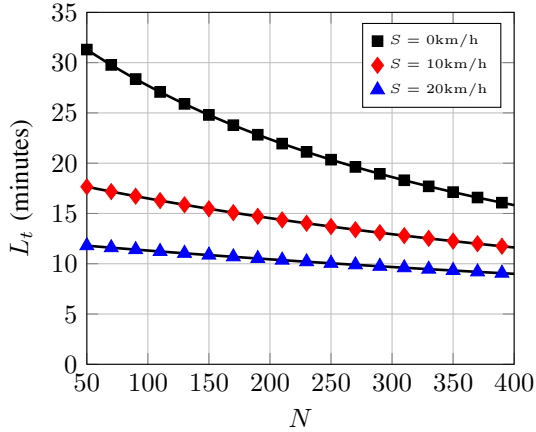
\begin{figure}
	\centering
	\begin{tikzpicture}
	\begin{axis}[
	width=0.82\linewidth,
	xlabel = {$N$},
	ylabel = {$L_t$ (minutes)},
	xmin = 50,xmax = 400,
	ymin = 0,
	ymax = 35,
	ytick = {0,5,...,35},
	xtick = {0,50,...,450},
	grid = major,
      legend style = {font = \tiny},
	legend cell align = {left},
	legend pos = north east
	]
	\addplot[
	black,
      only marks,
	mark=square*,
	mark repeat = 20,
	mark size = 2,
	]
	table {dat_new/lifetime_0_par.dat};
	\addlegendentry{$S= 0 \mathrm{km}/\mathrm{h}$}
	\addplot[
	red,
      only marks,
	mark=diamond*,
	mark repeat = 20,
	mark size = 3,
	]
	table {dat_new/lifetime_10_par.dat};
	\addlegendentry{$S= 10 \mathrm{km}/\mathrm{h}$}
	\addplot[
	blue,
      only marks,
	mark=triangle*,
	mark repeat = 20,
	mark size = 3,
	]
	table {dat_new/lifetime_20_par.dat};
	\addlegendentry{$S= 20 \mathrm{km}/\mathrm{h}$}
	\addplot[
	black,
       no marks,
	line width = 1pt,
	style = solid,
	]
	table {dat_new/lifetime_0_par.dat};
	%\addlegendentry{Analytical}
	\addplot[
	black,
      no marks,
	line width = 1pt,
	style = solid,
	]
	table {dat_new/lifetime_10_par.dat};
	\addplot[
	black,
      no marks,
	line width = 1pt,
	style = solid,
	]
	table {dat_new/lifetime_20_par.dat};
	\end{axis}
	\end{tikzpicture}
	\caption{UAV Lifetime versus $N$ for different UAV speeds.}
	\label{fig:lifetime}
\end{figure}

Regarding the UAV energy model, in Fig. \ref{fig:lifetime}, we present the effects of $N$ on the UAV lifetime, while considering three UAV speeds: i) hovering (0 km/h), ii) light speed (10 km/h), and iii) medium speed (20 km/h). It can be observed that, by increasing $N$, the battery's lifetime decreases due to the higher RIS weight and the corresponding increase in the UAV effort to lift the extra weight. In addition, the battery's lifetime decrease rate declines as the UAV's speed increases, indicating that the UAV speed plays a major role in the flight duration. Nevertheless, it is necessary to combine the UAV lifetime with the average throughput to characterize comprehensively the communication performance of our system.

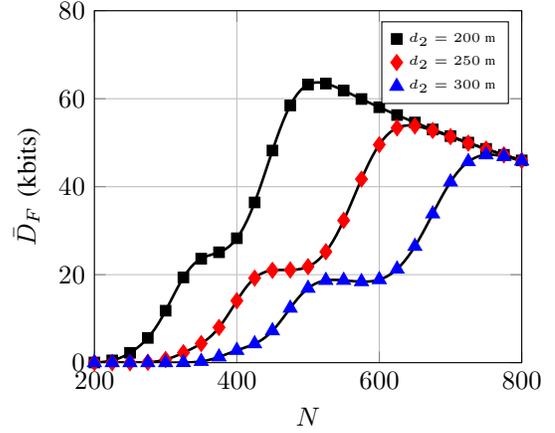
\begin{figure}
	\centering
	\begin{tikzpicture}
	\begin{axis}[
	width=0.82\linewidth,
	xlabel = {$N$},
	ylabel = {$\bar{D}_F$ (kbits)},
	xmin = 200,xmax = 800,
	ymin = 0,
	ymax = 80,
	ytick = {0,20,...,80},
	xtick = {0,200,...,800},
	grid = major,
      legend style = {font = \tiny},
	legend cell align = {left},
	legend pos = north east
	]
	\addplot[
	black,
      only marks,
	mark=square*,
	mark repeat = 5,
	mark size = 2,
	]
	table {data_new/AFD_200.dat};
	\addlegendentry{$d_2 = 200$ m}
	\addplot[
	red,
      only marks,
	mark=diamond*,
	mark repeat = 5,
	mark size = 3,
	]
	table {data_new/AFD_250.dat};
	\addlegendentry{$d_2 = 250$ m}
	\addplot[
	blue,
      only marks,
	mark=triangle*,
	mark repeat = 5,
	mark size = 3,
	]
	table {data_new/AFD_300.dat};
	\addlegendentry{$d_2 = 300$ m}
	\addplot[
	black,
       no marks,
	line width = 1pt,
	style = solid,
	]
	table {data_new/AFD_200.dat};
	%\addlegendentry{Analytical}
	\addplot[
	black,
      no marks,
	line width = 1pt,
	style = solid,
	]
	table {data_new/AFD_250.dat};
	\addplot[
	black,
      no marks,
	line width = 1pt,
	style = solid,
	]
	table {data_new/AFD_300.dat};
	\end{axis}
	\end{tikzpicture}
	\caption{$\bar{D}_F$ versus $N$ for different UAV-AP distances.}
	\label{fig:afd}
\end{figure}

To that end, in Fig. \ref{fig:afd}, we depict the average data per flight $\bar{D}_F$, which provides insights into the capabilities of the UAV-RIS system for the data collection. This figure shows the average collected data by the AP in kbits until the UAV has to return for recharging versus $N$ for different UAV-AP distances. As it can be observed, there exists an optimal number of reflecting elements that maximizes the performance of $\bar{D}_F$. Furthermore, the optimal RIS size increases as the UAV-mounted RIS moves away from the AP, indicating that for every sensor cluster, there exists a unique RIS that optimizes the data collection procedure. Hence, it is important for the network designer to adjust the RIS according to the needs and the distance of the IoT network.

\begin{figure}
	\centering
	\begin{tikzpicture}
	\begin{axis}[
	width=0.82\linewidth,
	xlabel = {$N$},
	ylabel = {$\bar{D}_F$ (kbits)},
	xmin = 200,xmax = 800,
	ymin = 0,
	ymax = 80,
	ytick = {0,20,...,100},
	xtick = {0,200,...,800},
	grid = major,
      legend style = {font = \tiny},
	legend cell align = {left},
	legend pos = north west
	]
	\addplot[
	black,
      only marks,
	mark=square*,
	mark repeat = 5,
	mark size = 2,
	]
	table {data_new/ADF_200_3.dat};
	\addlegendentry{$R = 20$ m, $L=3$}
	\addplot[
	blue,
      only marks,
	mark=triangle*,
	mark repeat = 5,
	mark size = 3,
	]
	table {data_new/ADF_200_1.dat};
	\addlegendentry{$R = 20$ m, $L=1$}
	\addplot[
	red,
      only marks,
	mark=diamond*,
	mark repeat = 5,
	mark size = 3,
	]
	table {data_new/ADF_200_3_60.dat};
	\addlegendentry{$R = 60$ m, $L=3$}
	\addplot[
	cyan,
      only marks,
	mark=*,
	mark repeat = 5,
	mark size = 2,
	]
	table {data_new/ADF_200_1_60.dat};
	\addlegendentry{$R = 60$ m, $L=1$}
	\addplot[
	black,
       no marks,
	line width = 1pt,
	style = solid,
	]
	table {data_new/ADF_200_3.dat};
	%\addlegendentry{Analytical}
	\addplot[
	black,
       no marks,
	line width = 1pt,
	style = solid,
	]
	table {data_new/ADF_200_1.dat};
	%\addlegendentry{Analytical}
	\addplot[
	black,
      no marks,
	line width = 1pt,
	style = solid,
	]
	table {data_new/ADF_200_3_60.dat};
	\addplot[
	black,
      no marks,
	line width = 1pt,
	style = solid,
	]
	table {data_new/ADF_200_1_60.dat};
	\end{axis}
	\end{tikzpicture}
	\caption{$\bar{D}_F$ versus $N$ for different radii $R$ and $L$ }
	\label{fig:afd_L}
\end{figure}
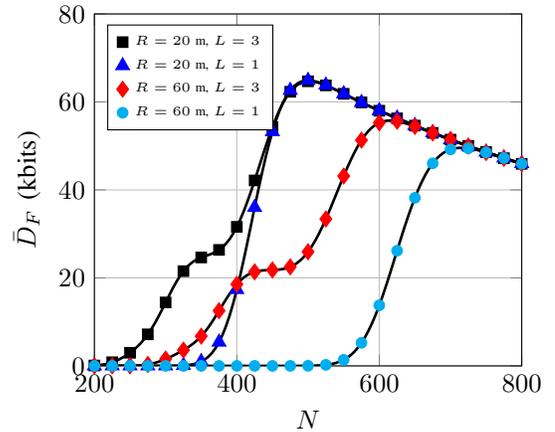

Finally, Fig. \ref{fig:afd_L} illustrates the impact of: i) the CC-based MAC protocol ($L=3$), and ii) the pure slotted ALOHA protocol ($L=1$), on the average data per flight versus $N$. The results are given for two different circular areas for the case where the UAV-AP distance is equal to 200m. As it can be observed, the utilization of the proposed MAC protocol can improve the data collection procedure even for RIS with fewer reflecting elements. In addition, by increasing the radius $R$, the utilization of the proposed MAC protocol can reduce the optimal number of reflecting elements, while it increases the maximum average data per flight compared with the pure slotted ALOHA case. Therefore, as the radius $R$ increases, the optimal RIS size that maximizes $\bar{D}_F$ changes, indicating that for larger circular areas where the sensors are located within, the UAV-mounted RIS should increase.

\section{Conclusions}\label{conc}
In this paper, we have investigated the data collection performance of a UAV-RIS synergetic system that serves over a remote area that has no direct link with the AP. To handle the communication of multiple sensors, we have proposed a novel MAC protocol based on slotted ALOHA and Code Combining. Furthermore, we have devised an energy model that takes into consideration the UAV and the RIS weight as well as the UAV's velocity and environmental conditions. In our results, we have characterized the performance of our model by analyzing the average throughput and the average data per flight, where by analyzing this novel metric, we prove that there exists a unique number of reflecting elements that optimizes the data collection procedure for a specific area which may change depending on the used MAC protocol. Therefore, we have shown that increasing the number of reflecting elements, i.e., the RIS size may lead to deteriorated data collection, which indicates the importance of proper RIS selection.

\bibliographystyle{IEEEtran}
\bibliography{Bibliography}

\begin{thebibliography}{10}
\providecommand{\url}[1]{#1}
\csname url@samestyle\endcsname
\providecommand{\newblock}{\relax}
\providecommand{\bibinfo}[2]{#2}
\providecommand{\BIBentrySTDinterwordspacing}{\spaceskip=0pt\relax}
\providecommand{\BIBentryALTinterwordstretchfactor}{4}
\providecommand{\BIBentryALTinterwordspacing}{\spaceskip=\fontdimen2\font plus
\BIBentryALTinterwordstretchfactor\fontdimen3\font minus
  \fontdimen4\font\relax}
\providecommand{\BIBforeignlanguage}[2]{{%
\expandafter\ifx\csname l@#1\endcsname\relax
\typeout{** WARNING: IEEEtran.bst: No hyphenation pattern has been}%
\typeout{** loaded for the language `#1'. Using the pattern for}%
\typeout{** the default language instead.}%
\else
\language=\csname l@#1\endcsname
\fi
#2}}
\providecommand{\BIBdecl}{\relax}
\BIBdecl

\bibitem{Guo}
F.~Guo, F.~R. Yu, H.~Zhang, X.~Li, H.~Ji, and V.~C.~M. Leung, ``{Enabling
  Massive IoT Toward 6G: A Comprehensive Survey},'' \emph{IEEE Internet Things
  J.}, vol.~8, no.~15, pp. 11\,891--11\,915, 2021.

\bibitem{6G}
Z.~Zhang, Y.~Xiao, Z.~Ma, M.~Xiao, Z.~Ding, X.~Lei, G.~K. Karagiannidis, and
  P.~Fan, ``{6G Wireless Networks: Vision, Requirements, Architecture, and Key
  Technologies},'' \emph{IEEE Veh. Technol. Mag.}, vol.~14, no.~3, pp. 28--41,
  2019.

\bibitem{eeiot}
N.~Kaur and S.~K. Sood, ``{An Energy-Efficient Architecture for the Internet of
  Things (IoT)},'' \emph{IEEE Syst J .}, vol.~11, no.~2, pp. 796--805, 2017.

\bibitem{uavkarag}
P.~S. Bithas, V.~Nikolaidis, A.~G. Kanatas, and G.~K. Karagiannidis,
  ``{UAV-to-Ground Communications: Channel Modeling and UAV Selection},''
  \emph{IEEE Trans. Commun.}, vol.~68, no.~8, pp. 5135--5144, 2020.

\bibitem{akis}
P.-V. Mekikis and A.~Antonopoulos, ``{Breaking the Boundaries of Aerial
  Networks with Charging Stations},'' in \emph{ICC 2019}, 2019, pp. 1--6.

\bibitem{wuUAV}
Q.~Wu, J.~Xu, Y.~Zeng, D.~W.~K. Ng, N.~Al-Dhahir, R.~Schober, and A.~L.
  Swindlehurst, ``{A Comprehensive Overview on 5G-and-Beyond Networks With
  UAVs: From Communications to Sensing and Intelligence},'' \emph{IEEE J. Sel.
  Areas Commun.}, vol.~39, no.~10, pp. 2912--2945, 2021.

\bibitem{hossain1}
T.~Shafique, H.~Tabassum, and E.~Hossain, ``{End-to-End Energy-Efficiency and
  Reliability of UAV-Assisted Wireless Data Ferrying},'' \emph{IEEE Trans.
  Commun.}, vol.~68, no.~3, pp. 1822--1837, 2020.

\bibitem{eezhang}
Y.~Zeng and R.~Zhang, ``{Energy-Efficient UAV Communication With Trajectory
  Optimization},'' \emph{IEEE Trans. Wireless Commun.}, vol.~16, no.~6, pp.
  3747--3760, 2017.

\bibitem{liaskos1}
C.~Liaskos, S.~Nie, A.~Tsioliaridou, A.~Pitsillides, S.~Ioannidis, and
  I.~Akyildiz, ``{A New Wireless Communication Paradigm through
  Software-Controlled Metasurfaces},'' \emph{IEEE Commun. Mag.}, vol.~56,
  no.~9, pp. 162--169, 2018.

\bibitem{eealex}
C.~Huang, A.~Zappone, G.~C. Alexandropoulos, M.~Debbah, and C.~Yuen,
  ``{Reconfigurable Intelligent Surfaces for Energy Efficiency in Wireless
  Communication},'' \emph{IEEE Trans. Wireless Commun.}, vol.~18, no.~8, pp.
  4157--4170, 2019.

\bibitem{liaskos2}
C.~Liaskos, L.~Mamatas, A.~Pourdamghani, A.~Tsioliaridou, S.~Ioannidis,
  A.~Pitsillides, S.~Schmid, and I.~F. Akyildiz, ``{Software-Defined
  Reconfigurable Intelligent Surfaces: From Theory to End-to-End
  Implementation},'' \emph{Proc. IEEE}, pp. 1--28, 2022.

\bibitem{RRS}
S.~A. Tegos, D.~Tyrovolas, P.~D. Diamantoulakis, C.~K. Liaskos, and G.~K.
  Karagiannidis, ``{On the Distribution of the Sum of Double-Nakagami-m Random
  Vectors and Application in Randomly Reconfigurable Surfaces},'' \emph{IEEE
  Trans. Veh. Technol.}, vol.~71, no.~7, pp. 7297--7307, 2022.

\bibitem{casc}
D.~Tyrovolas, S.~A. Tegos, E.~C. Dimitriadou-Panidou, P.~D. Diamantoulakis,
  C.~K. Liaskos, and G.~K. Karagiannidis, ``{Performance Analysis of Cascaded
  Reconfigurable Intelligent Surface Networks},'' \emph{IEEE Wireless Commun.
  Lett.}, pp. 1--1, 2022.

\bibitem{direnzo1}
E.~Basar, M.~Di~Renzo, J.~De~Rosny, M.~Debbah, M.-S. Alouini, and R.~Zhang,
  ``{Wireless Communications Through Reconfigurable Intelligent Surfaces},''
  \emph{IEEE Access}, vol.~7, pp. 116\,753--116\,773, 2019.

\bibitem{boulogeorgos}
A.-A. A.~Boulogeorgos, A.~Alexiou, and M.~D. Renzo, ``{Outage performance
  analysis of RIS-assisted UAV wireless systems under disorientation and
  misalignment},'' \emph{IEEE Trans. Veh. Technol.}, pp. 1--16, 2022.

\bibitem{synergetic}
D.~Tyrovolas, S.~A. Tegos, P.~D. Diamantoulakis, and G.~K. Karagiannidis,
  ``{Synergetic UAV-RIS Communication With Highly Directional Transmission},''
  \emph{IEEE Wireless Commun. Lett.}, vol.~11, no.~3, pp. 583--587, 2022.

\bibitem{uavdirenzo}
S.~Li, B.~Duo, X.~Yuan, Y.-C. Liang, and M.~Di~Renzo, ``{Reconfigurable
  Intelligent Surface Assisted UAV Communication: Joint Trajectory Design and
  Passive Beamforming},'' \emph{IEEE Wireless Commun. Lett.}, vol.~9, no.~5,
  pp. 716--720, 2020.

\bibitem{secrecyuav}
H.~Long, M.~Chen, Z.~Yang, Z.~Li, B.~Wang, X.~Yun, and M.~Shikh-Bahaei,
  ``{Joint Trajectory and Passive Beamforming Design for Secure UAV Networks
  with RIS},'' in \emph{2020 IEEE Globecom Workshops}, 2020, pp. 1--6.

\bibitem{kaddoum}
A.~Ranjha and G.~Kaddoum, ``{URLLC Facilitated by Mobile UAV Relay and RIS: A
  Joint Design of Passive Beamforming, Blocklength, and UAV Positioning},''
  \emph{IEEE Internet Things J.}, vol.~8, no.~6, pp. 4618--4627, 2021.

\bibitem{trungpoor}
K.~K. Nguyen, A.~Masaracchia, V.~Sharma, H.~V. Poor, and T.~Q. Duong,
  ``{RIS-assisted UAV Communications for IoT with Wireless Power Transfer Using
  Deep Reinforcement Learning},'' \emph{IEEE J. Sel. Top. Signal Process.}, pp.
  1--1, 2022.

\bibitem{timely}
\BIBentryALTinterwordspacing
A.~Al-Hilo, M.~Samir, M.~Elhattab, C.~Assi, and S.~Sharafeddine,
  ``{RIS-Assisted UAV for Timely Data Collection in IoT Networks},'' 2021.
  [Online]. Available: \url{https://arxiv.org/abs/2103.17162}
\BIBentrySTDinterwordspacing

\bibitem{haas}
T.~N. Do, G.~Kaddoum, T.~L. Nguyen, D.~B. da~Costa, and Z.~J. Haas, ``{Aerial
  Reconfigurable Intelligent Surface-Aided Wireless Communication Systems},''
  in \emph{2021 IEEE 32nd Annual International Symposium on Personal, Indoor
  and Mobile Radio Communications (PIMRC)}, 2021, pp. 525--530.

\bibitem{trung}
Y.~Li, C.~Yin, T.~Do-Duy, A.~Masaracchia, and T.~Q. Duong, ``{Aerial
  Reconfigurable Intelligent Surface-Enabled URLLC UAV Systems},'' \emph{IEEE
  Access}, vol.~9, pp. 140\,248--140\,257, 2021.

\bibitem{chatzinotas}
S.~Solanki, S.~Gautam, S.~K. Sharma, and S.~Chatzinotas, ``{Ambient Backscatter
  Assisted Co-Existence in Aerial-IRS Wireless Networks},'' \emph{IEEE Open J.
  Commun. Soc.}, vol.~3, pp. 608--621, 2022.

\bibitem{bothsides}
C.~You, B.~Zheng, W.~Mei, and R.~Zhang, ``{How to Deploy Intelligent Reflecting
  Surfaces in Wireless Network: BS-Side, User-Side, or Both Sides?}''
  \emph{J.Commn.Net}, vol.~7, no.~1, pp. 1--10, 2022.

\bibitem{qingqing}
Q.~Wu and R.~Zhang, ``{Intelligent Reflecting Surface Enhanced Wireless Network
  via Joint Active and Passive Beamforming},'' \emph{IEEE Trans. Wireless
  Commun.}, vol.~18, no.~11, pp. 5394--5409, 2019.

\bibitem{alouini1}
M.~Al-Jarrah, A.~Al-Dweik, E.~Alsusa, Y.~Iraqi, and M.-S. Alouini, ``{On the
  Performance of IRS-Assisted Multi-Layer UAV Communications With Imperfect
  Phase Compensation},'' \emph{IEEE Trans. Commun.}, vol.~69, no.~12, pp.
  8551--8568, 2021.

\bibitem{justin}
M.-A. Badiu and J.~P. Coon, ``{Communication Through a Large Reflecting Surface
  With Phase Errors},'' \emph{IEEE Wireless Commun. Lett.}, vol.~9, no.~2, pp.
  184--188, 2020.

\bibitem{Mardia2009}
K.~V. Mardia and P.~E. Jupp, \emph{Directional Statistics}.\hskip 1em plus
  0.5em minus 0.4em\relax John Wiley \& Sons, 2009, vol. 494.

\bibitem{SA}
S.~A. Tegos, P.~D. Diamantoulakis, A.~S. Lioumpas, P.~G. Sarigiannidis, and
  G.~K. Karagiannidis, ``{Slotted ALOHA With NOMA for the Next Generation
  IoT},'' \emph{IEEE Trans. Commun.}, vol.~68, no.~10, pp. 6289--6301, 2020.

\bibitem{alouiniCC}
A.~Chelli, E.~Zedini, M.-S. Alouini, M.~Pätzold, and I.~Balasingham,
  ``{Throughput and Delay Analysis of HARQ With Code Combining Over Double
  Rayleigh Fading Channels},'' \emph{IEEE Trans. Veh. Technol.}, vol.~67,
  no.~5, pp. 4233--4247, 2018.

\bibitem{motor}
\BIBentryALTinterwordspacing
T-MOTOR, ``{Technical Specifications},'' \emph{Datasheet}. [Online]. Available:
  \url{https://store.tmotor.com/goods.php?id=699}
\BIBentrySTDinterwordspacing

\bibitem{nature}
\BIBentryALTinterwordspacing
J.~K. Stolaroff, C.~Samaras, E.~R. O'Neill, A.~Lubers, A.~S. Mitchell, and
  D.~Ceperley, ``Energy use and life cycle greenhouse gas emissions of drones
  for commercial package delivery,'' \emph{Nature Communications}, vol.~9,
  no.~1, 2 2018. [Online]. Available: \url{https://www.osti.gov/biblio/1440731}
\BIBentrySTDinterwordspacing

\bibitem{vpapanikk}
V.~K. Papanikolaou, G.~K. Karagiannidis, N.~A. Mitsiou, and P.~D.
  Diamantoulakis, ``{Closed-Form Analysis for NOMA With Randomly Deployed Users
  in Generalized Fading},'' \emph{IEEE Wireless Commun. Lett.}, vol.~9, no.~8,
  pp. 1253--1257, 2020.

\bibitem{Coelho1998}
C.~A. Coelho, ``{The generalized integer Gamma distribution—a basis for
  distributions in multivariate statistics},'' \emph{Journal of Multivariate
  Analysis}, vol.~64, no.~1, pp. 86--102, 1998.

\end{thebibliography}

\end{document}